\documentclass[12pt]{article}
\usepackage{amsfonts}
\usepackage{latexsym}
\usepackage{amsmath}
\usepackage{amssymb}
\usepackage{amssymb}

\newcommand{\newsection}{    
\setcounter{equation}{0}\section}
\def\appendix#1{\addtocounter{section}{1}\setcounter{equation}{0}
\renewcommand{\thesection}{\Alph{section}}
\section*{Appendix \thesection\protect\indent \parbox[t]{11.15cm}{#1}}
\addcontentsline{toc}{section}{Appendix \thesection\ \ \ #1}}

\newcommand{\be}{\begin{eqnarray}}
\newcommand{\ee}{\end{eqnarray}}
\newcommand{\bea}{\begin{eqnarray}}
\newcommand{\eea}{\end{eqnarray}}
\newcommand{\ba}{\begin{array}}
\newcommand{\ea}{\end{array}}

\newenvironment{frcseries}{\fontfamily{frc}\selectfont}{}

\def \la {\label}

\def\a{\alpha}
\def\b{\beta}
\def\l{\lambda}

\def\e{\epsilon}

\font\mybb=msbm10 at 11pt

\def\bb#1{\hbox{\mybb#1}}

\def\bR {\bb{R}}

\begin{document}
\begin{titlepage}
\begin{center}
\vspace*{-1.0cm}

\vspace{2.0cm} {\Large \bf  New potentials for conformal  mechanics} \\[.2cm]

\vspace{1.5cm}
 {\large    G. Papadopoulos}

\vspace{0.5cm}


\vspace{0.5cm}
${}$ Department of Mathematics\\
King's College London\\
Strand\\
London WC2R 2LS, UK\\

\vspace{0.5cm}

\end{center}

\vskip 1.5 cm
\begin{abstract}
We find  under some mild  assumptions that the most general potential of 1-dimensional conformal systems with time independent couplings is expressed
as $V=V_0+V_1$,  where $V_0$  is a homogeneous function with respect to a homothetic motion in configuration space and $V_1$ is determined from an
equation with source a homothetic potential.
Such systems admit at most an $SL(2,\bR)$ conformal symmetry which,  depending on the couplings,
is embedded in   $\mathrm{Diff}(\bR)$ in three different ways. In one case, $SL(2,\bR)$ is also embedded in $\mathrm{Diff}(S^1)$.
 Examples of such models  include those with potential   $V=\alpha x^2+\beta x^{-2}$ for arbitrary
couplings $\a$ and $\b$, the Calogero models with harmonic oscillator couplings and
non-linear models with suitable metrics and potentials.  In addition, we give the conditions on the couplings for a class of gauge
 theories to admit a $SL(2,\bR)$  conformal symmetry. We present examples of such systems with general gauge groups and
 global symmetries that include the isometries of $AdS_2\times S^3$  and
$AdS_2\times S^3\times S^3$ which arise as backgrounds in $AdS_2/CFT_1$.

\end{abstract}

\end{titlepage}


\setcounter{section}{0}
\setcounter{subsection}{0}
\setcounter{equation}{0}
\newsection{Introduction}

It has been known for sometime that 1-dimensional models with potential $V=\beta x^{-2}$ are conformally invariant \cite{jackiwc, fubini}.  de Alfaro, Fubini and Furlan  (DFF) 
explored the
$SL(2,\bR)$ conformal symmetry of this theory and   noticed that the Hamiltonian operator  does not have a ground state \cite{fubini}.  To overcome this
problem,  they suggested to choose the eigenstates of
\bea
{\cal O}={p^2\over2}+\a x^2+\b x^{-2}~,
\la{op1}
\eea
as a basis in the Hilbert space. ${\cal O}$ is not the Hamiltonian operator, but a  linear combination  of conserved  charges associated with the $SL(2,\bR)$ conformal symmetry of the theory.  Choosing suitably the coupling constants  $\a, \b$ this
operator  exhibits a ground state  and discrete
  energy spectrum.
 As a result  the DFF formulation of the theory has been widely accepted in the literature. However, although a Hilbert space
 has been defined for the theory, the Hamiltonian operator is not diagonal
  in the chosen basis   and so the energy levels of the theory cannot be identified.  There have been many generalizations of the $V=\beta x^{-2}$
 model, see eg \cite{jackiw}-\cite{wyllard}, including the construction of non-linear theories\footnote{With the term  ``linear theories'' we mean those  for which the configuration space is $\bR^n$ equipped with the Euclidean metric
  but they  may exhibit a non-trivial potential. ``Non-linear theories'' are those with curved configuration space. }  \cite{michelson, papadopoulos},
which  exhibit similar properties, see also reviews \cite{michelsonb, olaf} and references within. The DFF treatment of the
theory and its generalizations have found widespread applications in the description of near
horizon black hole  dynamics \cite{claus, jade, gibbons, galajinskya} and in the understanding  of
black hole moduli spaces \cite{gibbonsb, shiraishi, gps, micha, gutpap}.

Another application of conformal mechanics is in the context of $AdS_2/CFT_1$ correspondence \cite{maldacena}, and for further exploration see eg \cite{sen, santos} . It is expected that string theory or M-theory on a $AdS_2\times X$ background is dual to a conformal theory on the boundary. After analytic continuation the Lorentzian boundary of $AdS_2$, which is two copies of $\bR$,
is mapped  to a circle, see eg \cite{sen}. In the Euclidean regime,  the associated dual theory should be a conformal theory defined on the circle. As we shall demonstrate, there are
such conformal theories but they are based on different potentials\footnote{The $SL(2,\bR)$ conformal symmetry of the $V=-\b x^{-2}$ model acts with fractional linear time
re-parameterizations and it cannot be embedded in $\mathrm{Diff}(S^1)$. We allow for diffeomorphisms with some discontinuities. }
 from $V=-\b x^{-2}$.

In this paper, we investigate the conformal properties of theories with Lagrangian
 \bea
 L={1\over2}\, g_{ij}\, \dot q^i \dot q^j-V~,
 \la{actiona}
 \eea
 where $g$ is a metric on the configurations space, $V$ is a potential  and $\dot q$ is the time derivative of the position.
 The conditions required for such theories to be invariant under the conformal transformations (\ref{transa})  have been stated in (\ref{cona}).
 Assuming that the configuration space of these theories admits
  a homothetic vector field $Z$
 associated with a homothetic potential $h$, the conditions for conformal invariance (\ref{cona}) can be solved. The potential of the theory can be written as
 \bea
 V=V_0+V_1~,
 \la{pot}
 \eea
 where $V_0$ is a homogeneous function with respect to the homothetic motion $Z$ and $V_1$ obeys the inhomogeneous
 equation (\ref{v1}) which has source the homothetic potential $h$. The dimension of the conformal group of these models is at most three and one of the generators is time translations.
 This is because the parameter of the transformation obeys a third order equation (\ref{pareqn}).
 The maximal conformal group is $SL(2,\bR)$ and it is embedded in $\mathrm{Diff}(\bR)$ in three different ways generating the vector fields
 \bea
&(i)&~~\partial_t,~~~ t\partial_t,~~~t^2\partial_t;
\cr
&(ii)&~~\partial_t,~~~ \cosh{(\omega t)}\partial_t,~~~ \sinh({\omega t})\partial_t;
\cr
&(iii)&~~\partial_t,~~~ \cos(\omega t)\partial_t,~~~ \sin(\omega t)\partial_t;
\la{vect}
\eea
for some $\omega$ related to the couplings.

The first $SL(2,\bR)$ embedding (i) in (\ref{vect}) is realized for the models with $V_1=0$. These class of models has a homogeneous potential $V_0$
and includes the DFF model,  and its linear and non-linear generalizations
\cite{michelson, papadopoulos}.
Furthermore, if $V_1\not=0$, the $SL(2,\bR)$ conformal group is embedded in $\mathrm{Diff}(\bR)$ generating the  vector fields (ii) or (iii).
These are Newton-Hooke transformations and the two cases are distinguished by the sign of the inhomogeneous
term in the equation (\ref{v1}) which determines $V_1$.
The models with conformal transformation  (ii) and (iii)  are related by a naive analytic continuation,
and the $SL(2,\bR)$ group in the latter case can be embedded in $\mathrm{Diff}(S^1)$.

The class of conformal models with  conformal symmetry  (ii) and (iii) in (\ref{vect}) includes those with potential \cite{jackiw}
\bea
V= \alpha x^2+\beta x^{-2}~,
\la{pota}
\eea
where $V_0=\beta x^{-2}$ and $V_1=\alpha x^2$. 
For $\a<0$ the conformal group generates the vector fields (ii) in (\ref{vect}), while for  $\a>0$ the conformal group generates the vector field (iii).
There are also several multi-particle models which exhibit type (ii) and (iii) in (\ref{vect}) conformal symmetry.  Such systems include the Calogero model  with harmonic oscillator couplings of equal frequency \cite{calogero}, and  the multi-particle  linear models of \cite{galajinsky} for which $V_0$ satisfies additional symmetries. We shall present some additional linear and non-linear systems with (ii) and (iii) conformal symmetries.
 Observe that the theories with $\a,\b>0$ in (\ref{pota}) have a ground state and discrete
 energy spectrum, and so there is no need to choose another operator different from the Hamiltonian to give a basis in the Hilbert space of the theory. This also applies
 to several other models in this class.

One result which follows from the general analysis of this paper is that the most general  linear conformal  model admits a potential (\ref{pot}), where $V_0$ is a homogeneous function of the positions $q$ of degree -2 and $V_1=\a |q|^2$. This rigidity result
is based on the uniqueness of homothetic motions in flat space associated with a homothetic potential. The homothetic motion is the homogeneous scaling of
all coordinates, $q^i\rightarrow \ell q^i$. These models admit an $SL(2,\bR)$  conformal symmetry  generated by the vector
fields (ii) and (iii) in (\ref{vect}) and depending  on whether $\a<0$ or $\a>0$, respectively.

More recently, conformal models in one dimension have been investigated which apart from scalar fields contain also vectors \cite{olafb}. So far such theories
have been based on gauging models with homogeneous potentials. We shall demonstrate that such models can be generalized to include potentials of the type (\ref{pota}).
In particular, we derive the conditions (\ref{conb})  for gauged non-linear sigma models with Lagrangian (\ref{actionb})
  to admit a conformal symmetry, and determine the equations that restrict the potentials.
  We find that for a large class of such conformal theories the potential can be written as in (\ref{pot}), where both $V_0$ and $V_1$ must also be gauge invariant. In addition,
we give some examples which include conformal models with a general gauged group and global symmetries. Some of these models exhibit the isometries of
$AdS_2\times S^3$ and $AdS_2\times S^3\times S^3$ backgrounds as global symmetries. A class of these models is solvable, and the Hamiltonian has a ground state and
discrete spectrum. A similar investigation of $SL(2,\bR)$ symmetries in the context of matrix models has been done
 in \cite{park} and the associated potentials have been identified.

This paper is organized as follows. In section 2, we derive  investigate the conditions for conformal invariance of non-linear 1-dimensional theories and derive the scalar potential (\ref{pot}).  In section 3, we give several examples of such models. In section 4, we derive the conditions on the  couplings
 gauged sigma models with a potential to admit conformal invariance, and give several examples. In section 5, we present our conclusions.

\newsection{Conformal models}

\subsection{Lagrangian}

Consider the Lagrangian (\ref{actiona}) of a sigma model
on a manifold $M$ with metric $g$ and with  a potential $V$. This describes either  the propagation of a non-relativistic particle in a curved manifold $M$ or a multi-particle
system with a non-trivial configuration space $M$.
One can assign mass dimensions such that $q$ is dimensionless  $[q]=0$ while $[t]=-1$. Thus $[L]=2$ provided
one takes the coupling $V$ terms to have dimension 2.  This   is not the most general Lagrangian that one can consider as a coupling with dimension 1 has not been included.
This will be done elsewhere \cite{new}.

\subsection{Conformal transformations}

All time re-parameterizations  $t'=u(t)$ are conformal transformations of the Euclidean metric on $\bR$ as $ds^2=(dt')^2= (\dot u)^2 dt^2$.  Therefore, one can choose
any of these transformations and demand that leave the action (\ref{actiona}) invariant. Apart from time translations\footnote{We have chosen the couplings $g$ and $V$
not to depend  explicitly on time. However, it is straightforward to carry out the analysis of this section for models with time-dependent couplings.}, such transformations will not leave the action
invariant unless there is a compensating additional transformation on the positions generated by a vector field $X$ on $M$ \cite{michelson}.  As a result, one considers the
infinitesimal transformations \cite{papadopoulos}
\bea
\delta q^i=-\e a(t) \dot q^i+\e X^i(t,q)~,
\la{transa}
\eea
where $\e$ is a small parameter. The first term in the transformation of $q$ is induced by the infinitesimal transformation $\delta t=\e a(t)$, where $a(t)$ is the vector field
on $\bR$ which generates the time re-parameterizations, while the second term containing $X$ is the compensating transformation  which
may  explicitly depend on $t$.

The conditions for the invariance of the action (\ref{actiona}), up to surface terms, under the transformations (\ref{transa})  are \cite{papadopoulos}
\bea
{\cal L}_X g_{ij}=\dot a g_{ij}~,~~~
\partial_tX^i g_{ij}=\partial_i f~,~~~
\dot a V+X^k \partial_k V=-\partial_t f
\la{cona}
\eea
where $f=f(t, q)$ is the contribution from the surface term,  and where $\partial_t $ denotes differentiation of the explicit dependence of $X$ and $f$ on $t$, ie
\bea
{d\over dt} f(q,t)= \partial_t f+ \dot q^i \partial_i f~.
\eea

The conserved charges associated with the above symmetries are
\bea
Q(a, X)= {a\over2} g_{ij} \dot q^i \dot q^j-g_{ij} \dot q^i X^j+a V+f~.
\eea
It can be easily shown that $Q(a, X)$ is conserved subject to field equations.
\subsection{Solution of conformal conditions and new models}

It is clear that the first condition in (\ref{cona}) implies that $X$ generates a family of homothetic transformations on $M$ which may depend on $t$.
Since all  $\mathrm{Diff}(\bR)$ are conformal transformations, the system can be invariant under any subgroup of  $\mathrm{Diff}(\bR)$. So, one should consider at most as many homothetic motions in $M$ as the dimension of the subgroup of conformal transformations.
However, in most examples of interest $M$ admits one homothetic motion generated by a vector field $Z$ which does not depend explicitly
on $t$
\bea
{\cal L}_Z g_{ij}=\ell  g_{ij}~,
\eea
where $\ell$ is a constant.
Then, the first condition can be solved by setting
\bea
X^i(t, q)= \ell^{-1} \dot a(t)\, Z^i(q)~.
\la{xaz}
\eea
Assuming that $Z$ arises from a homothetic potential, ie
\bea
Z^i g_{ij}=\partial_j h~,
\eea
where $h=h(q)$, $f$ can be chosen\footnote{We assume that $\ddot a\not=0$. If $\ddot a=0$, $Z$ does not have to be associated with a homothetic potential
and $V$ is a homogeneous function of the homothetic motion.  The models do not have a $SL(2,\bR)$ symmetry but rather are invariant under time translations
and scale transformations generated by the vector fields $\partial_t, t\partial_t$.}
\bea
f=\ell^{-1}\ddot a h~.
\eea
The last equation in (\ref{cona}) can now be rewritten as
\bea
\dot a (V+\ell^{-1}Z^k\partial_k V)=-\ell^{-1}\partial_t^3 a\, h~.
\eea
Since we are seeking to find potentials $V$ which solve the above equations and  do not depend explicitly on $t$, we have to take
\bea
\partial_t^3 a=\lambda \dot a~,
\la{pareqn}
\eea
where $\lambda$ is a constant.  Of course, if $\dot a=0$, there is no condition on  $V$  as the only symmetry of the action is time translations.
Thus, we take $\dot a\not=0$ and
as a result the equation which determines the potential is
\bea
V+\ell^{-1} Z^k\partial_k V=-\ell^{-1} \lambda h~.
\eea
The general solution for the potential can be written as in (\ref{pot}), ie $V= V_0+V_1$,
where $V_0$ is the most general  solution of the homogenous equation
\bea
V_0+\ell^{-1} Z^k\partial_k V_0=0~,
\la{v0}
\eea
and $V_1$ is a solution of
\bea
V_1+\ell^{-1} Z^k\partial_k V_1=-\ell^{-1} \lambda h~.
\la{v1}
\eea

Clearly, there are 3 cases to consider depending on whether $\lambda=0$, or $\lambda>0$ or $\lambda<0$.
In these three choices, the vector field $a$ is determined from (\ref{pareqn}) as follows.  For $\lambda=0$, one has
\bea
a=a_0+a_1 t+a_2 t^2~,
\la{l0}
\eea
where $a_0, a_1$ and $a_2$ are integration constants. For $\lambda=\omega^2$, one has
\bea
a=a_0+ b e^{\omega t}+ c e^{-\omega t}~,
\la{lp}
\eea
 and for $\lambda=-\omega^2$, one has
\bea
a=a_0+ b \cos(\omega t)+ c \sin(\omega t)~,
\la{ln}
\eea
where $a_0, b, c$ are integration constants. The new conformal models arise from the last two cases.

Before we proceed to investigate individual models, let as examine the algebra of these transformations. A basis in the space of vector fields of the
 infinitesimal transformations (\ref{l0}), (\ref{lp}) and (\ref{ln}) is given in (i), (ii) and (iii) of (\ref{vect}), respectively, with  $|\lambda|=\omega^2$. The group of transformations
 generated by (\ref{l0}),  (\ref{lp}) and (\ref{ln}) is $SL(2,\bR)$. However, $SL(2,\bR)$ is embedded into  $\mathrm{Diff}(\bR)$ in three different ways\footnote{In the (\ref{l0})
 case, $SL(2,\bR)$ acts with fractional linear transformations on $\bR$.}. The group of transformations generated by (\ref{ln}) is also embedded in the $\mathrm{Diff}(S^1)$ as the associated vector fields are periodic in $t$. The two cases (\ref{lp}) and (\ref{ln}) are related
 to each other by analytic continuation.

Substituting the above expressions of $X$ into the conserved charges and using the properties of the homothetic motion on $M$, one finds that
\bea
Q(a, Z)={a\over 2} g_{ij} \dot q^i \dot q^j- \dot a \ell^{-1} \partial_i h \dot q^i+ a (V_0+V_1)+ \ell^{-1}\ddot a h~.
\la{cha}
\eea
These can be easily computed explicitly in the examples described below.

\newsection{Examples}

\subsection{Conformal particle in flat space}

The most illuminating model is that of a single particle propagating on the real line.  Here we shall show that (\ref{pota}), which has been found previously in \cite{jackiw},  is the only potential 
consistent with conformal invariance. For this we shall take the Lagrangian
\bea
{\cal  L}= {1\over2} \dot x^2-V(x)~,
\eea
and we shall determine $V$ such that the action is conformally invariant.
For this consider the homothetic vector field
\bea
Z={1\over 2}  x\partial_x~,
\eea
on the configuration space. For this choice of $Z$, $\ell=1$. The homothetic potential in this case is
\bea
h={1\over 4} x^2~.
\eea
Then the equation (\ref{v0}) can be solved for $V_0$ to yield
\bea
V_0=\beta x^{-2}~,
\eea
for some constant $\beta$, which the potential of the DFF model. However, we have  seen that the potential $V$ also receives a contribution from $V_1$
which is determined in (\ref{v1}). The latter equation can be solved as
\bea
V_1=\a\, x^2~,~~~\a=-\lambda/8~.
\eea
Thus the most general potential $V=V_0+V_1$ of such conformal models is given in (\ref{pota}).

The  Hamiltonian of this class of conformal  models is given in (\ref{op1}). As it has already been mentioned
the associated Hamiltonian operator with $\a>0, \b\geq 0$ has a ground state and discrete spectrum.

\subsection{Conformal multi-particle systems}

Consider next the linear model of $N$ particles propagating in $\bR$ and interacting with a potential $V$. The Lagrangian of such a system is
\bea
{\cal L}={1\over 2} \sum_i^N (\dot x^i)^2- V(x^i)~.
\la{laction}
\eea
To find the potentials $V$ consistent with conformal invariance, consider the homothetic motion
\bea
Z={1\over2}\sum_{i=1}^N  x^i \partial_i~,
\la{homoz}
\eea
of $\bR^N$  configuration space. The homothetic potential in this case is
\bea
h= {|x|^2\over 4}~,~~~ |x|^2=\delta_{ij} x^i x^j~.
\eea
As it has been mentioned in the introduction, $Z$ in (\ref{homoz}) is the unique homothetic motion in $\bR^N$ associated with a homothetic potential\footnote{If the requirement
of the homothetic potential is removed, the scaling transformation (\ref{homoz}) can mix with other isometries, like $SO(N)$ rotations, to give rise to new homothetic motions.
 These can be used to construct invariant theories under subgroups of $SL(2,\bR)$ involving at most two generators.} up to an overall scale
which does not affect the form of the potential. After solving the conditions (\ref{v0}) and (\ref{v1}), one finds that the potential $V$ is
\bea
V= \a |x|^2+ V_0(x)~,~~~\a=-\lambda/8
\la{cal1}
\eea
and $V_0$ is a homogeneous function of degree $-2$
\bea
x^i\partial_i V_0=-2 V_0~.
\la{hom}
\eea
 (\ref{cal1}) is the most general potential of linear models.

Of course, there are many choices for $V_0$. A minimal choice for $V_0$ is $V_0=\b |x|^{-2}$. However, this is not unique. For example, one can also choose
\bea
V_0=\sum_{i\not= j}{\b_{ij}\over (x^i-x^j)^2}~.
\la{cal2}
\eea
The models with potentials $V$ given in (\ref{cal1}) and (\ref{cal2}) are the Calogero models with harmonic couplings of equal frequency.    Our results demonstrate that these models
 are conformally invariant. It is well-known that such models
with $\a>0$ and $\b\geq 0$ have a vacuum state and
 discrete energy spectrum \cite{calogero, mende}. Of course, there are many more potential functions $V_0$ which satisfy the homogeneity condition (\ref{hom})
above than those appearing in the Calogero models. The above models also include those presented in  \cite{galajinsky} where some additional symmetry assumptions
were made on the form of $V_0$ potential.

 To summarize, we have shown that all the above models admit either an $SL(2,\bR)$ conformal symmetry which is embedded in $\mathrm {Diff}(\bR)$
as in  (i),  (ii)  or (iii) of (\ref{vect}) depending on whether $\a=0$, $\a<0$  or $\a>0$, respectively.  The associated conserved charges can be computed by a direct substitution in (\ref{cha}).

\subsection{Particles propagating on cones}

So far, we have presented linear models as examples. For a non-linear example, consider particles propagating on a cone and interacting
with a potential $V$. The Lagrangian of such a system is
\bea
{\cal L}={1\over 2} \big( \dot r^2+ r^2 \gamma_{ij} \dot x^i \dot x^j\big)-V(r, x)~,
\eea
where $\gamma$ is the metric of the cone section which does not depend on the radial coordinate $r$ but it may
depend on the rest of the coordinates $x$.
The cone metric
\bea
ds^2=dr^2+ r^2 \gamma_{ij} dx^i dx^j~,
\eea
admits a homothetic motion generated by the vector field
\bea
Z={1\over2} r \partial_r~,
\eea
which homothetic potential
\bea
h={r^2\over 4}+ k(x)~,
\eea
where $k$ is an arbitrary function of $x$.
It is straightforward to show that the most general potential compatible with conformal symmetry is
\bea
V=\a r^2+ \b(x) r^{-2}+8 \a k(x)~,~~~\a=-\lambda/8~.
\eea
Again these models admit a $SL(2,\bR)$  conformal symmetry generating the vector fields (i),  (ii)  or
(iii) of (\ref{vect}) depending on whether  $\a=0$,  $\a<0$ or $\a>0$,  respectively.

\newsection{Conformal gauge theories in one dimension}

\subsection{Action}

Motivated by applications in $AdS/CFT$,  which typically
requires dual theories with a gauge symmetry,  and to enhance the class of 1-dimensional conformal systems, we shall also examine the conditions for a gauged sigma model to admit conformal invariance.
For this, we assume
 that  $M$ admits a group of isometries $G$, generating the vector fields $\xi$, which leave $V$ invariant.
 Gauging  the isometries of (\ref{actiona}), one finds the Lagrangian\footnote{This is not the most general Lagrangian of dimension 2 as couplings of dimension 1 have not
 been included.}
\bea
L={1\over2} g_{ij} \nabla_t q^i \nabla_t q^j-V~,
\la{actionb}
\eea
where
\bea
\nabla_t q^i= \dot q^i-A^a \xi_a^i~,~~~[\xi_a, \xi_b]=-f_{ab}{}^c \xi_c~,
\eea
 $A$ is the gauge potential and $f$ are the structure constants of $G$. We assign mass dimension to $A$ as $[A]=1$ so that $L$ has mass dimension 2.

The equations of motion of the theory are
\bea
g_{ij} D_t\nabla_t q^j+\partial_i V=0~,~~~\xi_{ia} \nabla_t q^i=0~,
\eea
where
\bea
D_t\nabla_t q^i=\partial_t\nabla_t q^i-A^a \partial_j \xi^i_a \nabla_tq^j+\Gamma^i_{jk} \nabla_t q^j \nabla_t q^k~.
\eea

Under certain conditions the gauge connection $A$ can be eliminated from the equations of motion leading to a theory
with dynamical  variables just the $q$'s. In particular notice that the second equation of motion can be rewritten as
\bea
\ell_{ab} A^b= \xi_{ia} \dot q^i
\eea
where $\ell_{ab}= g_{ij} \xi^i_a \xi^j_b$. If $\ell$ is invertible, then all $A$ can be eliminated. However, we shall not elaborate on this here.
Instead, we shall proceed to find the conditions such that the action (\ref{actionb}) is invariant under some conformal symmetries.

\subsection{Conformal and gauge  symmetries}

The action (\ref{actionb}) is invariant under the gauge transformations
\bea
\delta q^i=\eta^a \xi_a^i~,~~~ \delta A^a=\nabla_t \eta^a~,
\la{gauge}
\eea
where $\eta$ is the gauge infinitesimal parameter.

Next as in the un-gauged case, one expects that the transformations on $q$ and $A$, which induce the conformal symmetries of  the action (\ref{actionb}), to contain two parts. One
part is associated with  time re-parameterizations and an additional term which generates  compensating transformations on the configuration space.
As a result, we postulate the conformal transformations
\bea
\delta q^i&=&-\e a(t) \partial_t q^i+\e X^i(t, q, A)~,
\cr
\delta A^a&=& -\e \dot a A^a-\e a \dot A^a+\e W^a(t, q, A)~,
\eea
where the first term in the variation of $q$ and the first two terms in the variation of $A$ are the transformations
induced on $q$ and $A$ from the infinitesimal re-parameterization of $t$, $\delta t=\e a(t)$, and  the rest
 are the compensating transformations.

These transformation mix with the gauge transformations above. In particular, the coordinate transformation induced on $A$ by $a$ can be rewritten
as a gauge transformation with parameter $-a A^a$. Since the action is invariant under gauge transformations, this can be used to simplify
the conformal transformations as
\bea
\delta q^i&=&-\e a(t) \nabla_t q^i+\e X^i~,
\cr
\delta A^a&=& \e W^a~.
\la{conformal}
\eea
For the same reason $X$ and $Z$ are not uniquely defined. In particular $X$ and $W$ are defined up to terms $\ell^a \xi_a$ and $\nabla_t\ell^a$, respectively,
where $\ell=\ell(t, q,A)$.

Assuming that $X$ and $W$ do not depend on time derivatives of $q$, a straightforward computation reveals  that the conditions required for the invariance of the action, up
to surface terms, are
\bea
{\cal L}_Xg_{ij}=\dot a g_{ij}~,
\cr
g_{ij} \partial_tX^j+g_{ij} A^a [\xi_a, X]^j- g_{ij}\xi^j_b W^b=\partial_i f~,
\cr
\dot a V+X^k\partial_k V=-\partial_t f~,
\la{conb}
\eea
where $f=f(t,q)$ is the contribution from the surface term. $f$  is taken to be  gauge invariant, $\xi_a^i \partial_i f=0$. To find conformal models, one has to solve (\ref{conb}).

\subsection{Solution of conformal conditions}

Here, we shall not seek the most general solution to the conformal invariance conditions(\ref{conb}).  Instead, we shall take
\bea
[\xi_a, X]=0~,~~~ W^a=0~.
\la{com}
\eea
In this case, the above conditions (\ref{conb}) reduce to those of (\ref{cona}) but with the additional assumption that $f$ is gauge invariant.

To find solutions, we proceed as in section 2.3. The potential is given as $V=V_0+ V_1$, (\ref{pot}), with $V_0$ and $V_1$ determined by the
equations (\ref{v0}) and (\ref{v1}), respectively. There is an additional restriction here that the homothetic potential $h$ is
gauge invariant, $\xi_a^i\partial_i h=0$.

As in the systems without gauge symmetry, there are three cases to consider depending on whether $\l=0$, $\l>0$ or $\l<0$.
In all cases the conformal group is $SL(2, \bR)$ but it is embedded in three different ways into  $\mathrm{Diff}(\bR)$. The $\l>0$ and $\l<0$ models are related by
analytic continuation.

\subsection{Examples}

\subsubsection{Gauged nonlinear models on a cone}

Examples of non-linear gauge theories exhibiting conformal symmetry are those that describe the propagation of  particles on a cone. Assuming that
the cone section metric $\gamma$ admits a group of isometries generating the vector fields $\xi$, the Lagrangian of the theory can be written as
\bea
{\cal L}={1\over2} \big( \dot r^2+ r^2 \gamma_{ij} \nabla_t x^i \nabla_t x^j\big )- V(r,x)~,
\eea
where
\bea
\nabla_t x^i=\dot x^i- \xi_a^i A^a~.
\eea
The homothetic vector field is again given by $Z={1\over2} r\partial_r$ and commutes with the Killing vector fields $\xi_a$
satisfying the assumption (\ref{com}).

The rest of the analysis proceed as in the cone example in section 3.3 for the un-gauged model yielding a potential
\bea
V=\a\, r^2+\beta(x)\, r^{-2}+8 \a k(x)~,~~~\a=-\lambda/8~,
\la{potb}
\eea
where now $\b(x)$ and $k(x)$ are gauge invariant functions of the cone section, $\xi^i_a \partial_i \b=\xi^i_a \partial_i k=0$. The simplest explicit example is to consider the
flat cone $\bR^2$ and as the gauged symmetry the rotational symmetry. The potential of this model is given as in (\ref{potb}) with
$\b$ and $k$ constants.

\subsubsection{Gauge theories}

A large class of linear conformal models\footnote{These can also be thought of as special cases of the cone models above.} can be constructed beginning from some gauge group $G$ and some linear representation $D$ of its Lie algebra
$\mathfrak{g}$ on a vector space
$\mathcal{V}$. Suppose that $D$ leaves invariant a (constant) metric $g$ on $\mathcal{V}$. Then one can consider the Lagrangian
\bea
L={1\over2} g_{mn} \nabla_t x^m \nabla_t x^n- V(x)~,
\la{actionc}
\eea
where
\bea
 \nabla_t x^m= \dot x^m- A^a (D_a)^m{}_n x^n~.
 \eea
 To determine $V$ such that this  theory is conformal, observe that the metric admits a homothetic motion generated
 by the vector field
 \bea
 Z={1\over2}x^m \partial_m~.
 \eea
 Moreover, this commutes with the Killing vector fields
 \bea
 \xi_a= {1\over2} (D_a)^m{}_n x^n\partial_m~,
 \eea
 ie $[Z, \xi_a]=0$. As a consequence (\ref{com}) is satisfied. Furthermore, the homothetic potential of $Z$ is
 \bea
 h={1\over4} g_{mn} x^m x^n~.
 \eea
 Using this, the potential $V$ can be determined by solving (\ref{v0}) and (\ref{v1}) as
 \bea
 V=\a g_{mn} x^m x^n+ V_0~,~~~\a=-{\l\over 8}~,
 \eea
 and $V_0$ is a function of $x$ of homogeneous degree -2,
 \bea
 x^m\partial_m V_0=-2 V_0~,
 \eea
  which is also invariant under $G$. The minimal choice is
 \bea
 V_0={\b\over  g_{mn} x^m x^n}~.
 \la{minv0}
 \eea
However such a choice is not unique for general gauge groups and representations $D$. A similar potential has been derived
in the investigation of $SL(2,\bR)$ invariant matrix models in \cite{park}.

 Amongst these models, one can take as $D=\mathfrak{adj}\otimes I^k$, where $\mathfrak{adj}$ is the adjoint representation of a group $G$ and $I$ is the trivial representation. In such a case, the Lagrangian can be written as
 \bea
L={1\over2} g_{ab} \kappa_{ij} \nabla_t x^{ai} \nabla_t x^{bj}- V(x)
\eea
 where
 \bea
 \nabla_t x^{ai}= \dot x^{ai}- A^b f_{bc}{}^a x^{ci}~,
 \eea
 $g_{ab}$ is an invariant metric on the adjoint representation of $G$ and $\kappa$ a metric on the k-copies of the trivial representation.
 The potential in this case can be written as
 \bea
 V=\a g_{ab}\kappa_{ij} x^{ai} x^{bj}+ V_0~,~~~\a=-{\l\over 8}~,
 \la{potb}
 \eea
and $V_0$ is a function of $x$ of homogeneous degree -2 which is also invariant under $G$.  Now there are several options for $V_0$. For example,
$V_0$ can be any  homogeneous function of degree -2 expressed in terms of the gauge invariant functions like
\bea
m^{ij}=g_{ab} x^{ai} x^{bj}~,~~~m^{ijk}=f_{abc} x^{ai} x^{bj} x^{ck}~,
\eea
and many others which can be constructed from all the invariant tensors of $\mathfrak{g}$ under the action of the adjoint representation.
One example is a gauged Calogero model for which the potential is given in   (\ref{potb}) with
\bea
V_0=\sum_{i\not=j} {\beta_{ij} \over g_{ab} (x^{ai}- x^{aj}) (x^{bi}- x^{bj})}~.
\eea
Further restrictions can be put on the form of the  potential by requiring that the theory is invariant under the global symmetry $\times_iO(n_i)$ which leaves $\kappa$ invariant.
The above construction can also be done by replacing  $\mathfrak{adj}$ with another representation of the gauge group.

This class of conformal theories has all the bosonic symmetries required for the CFT duals of backgrounds like $AdS_2\times S^3$ or $AdS_2\times S^3 \times S^3$. In particular,
one can easily construct models with rigid symmetry $SL(2,\bR)\times SO(4)$, which is the isometry group of $AdS_2\times S^3$, and any gauge symmetry including $U(N)$, and similarly
there are models which exhibit the isometries of $AdS_2\times S^3 \times S^3$ backgrounds as symmetries. It is also worth remarking that the analytic continuation of a $\l>0$ theory
which exhibits $SL(2,\bR)$ conformal symmetry is equivalent to taking  $\lambda$ to $-\lambda$ and $V_0$ to $-V_0$ and  leads to a model with $SL(2,\bR)$ conformal invariance but now embedded in  $\mathrm{Diff}(S^1)$  as expected in the context of AdS${}_2$/CFT${}_1$.

The quantum theory of the model with action  (\ref{actionc}) can be easily described in the case that $V_0=0$ and $\a>0$.  The Hilbert space of these theories can be constructed starting from the Hilbert space   of $\mathrm{dim} D$ harmonic oscillators. Then, gauge invariance requires that one has to consider only those states which are invariant under the gauge group. The Hamiltonian operator has a ground state and the spectrum is discrete. However the details of the construction depend on the choice of gauge group and representation $D$.
If $V_0\not=0$,  the quantum theory depends on the choice of $V_0$.
 It is likely
that some of the properties of the $V_0=0$ models can be maintained in the presence of a large class of $V_0$ potentials as it happens for the Calogero models with harmonic oscillator couplings.

\newsection{Concluding remarks}

We have demonstrated that the potential $V$ of conformal mechanics models admitting a homothetic motion in configuration space can be expressed as a sum $V=V_0+V_1$,
 where $V_0$ is a homogeneous function of the  homothetic motion and $V_1$ is determined from an equation which has as a source the homothetic potential.
 Depending on the couplings, the maximal conformal group  $SL(2,\bR)$
  is embedded in $\mathrm{Diff}(\bR)$ in three different ways. Furthermore, one of these  can also be thought as an embedding of $SL(2,\bR)$ in
 in $\mathrm{Diff}(S^1)$. This is significant from the point of view of $AdS_2/CFT_1$ as the dual Euclidean theory must be defined
 on the boundary which is a circle.

 Examples of conformal 1-dimensional systems include  models with potential
  $V= \a x^2+ \b x^{-2}$ \cite{jackiw}. The $SL(2,\bR)$ conformal symmetry of this model  is embedded in $\mathrm{Diff}(\bR)$ in three different ways depending on whether $\a=0$,  $\a<0$ or $\a>0$,
  respectively.  Moreover if $\a>0$, $SL(2,\bR)$
can also be embedded in $\mathrm{Diff}(S^1)$.

We have described all 1-dimensional linear conformal  theories  described by the  Lagrangian (\ref{laction}).  The potential of all such models is
$V=\a |x|^2+ V_0$, where $V_0$ is a homogeneous of degree $-2$ function of the positions $x$.  This rigidity  result is
based on the  uniqueness of the homothetic motion in flat space associated with a homothetic potential and the analysis
in section 2. Examples of such theories include the Calogero models with harmonic oscillator couplings of equal
 frequency as well as the models given in \cite{galajinsky}. We have also presented examples of non-linear models.

It is clear  form the analysis of section 2  that if the configuration space of a system admits a single homothetic motion associated with a homothetic potential, then the vector field $a(t)\partial_t$ which generates the time
re-parameterizations obeys the third order equation (\ref{pareqn}). Because of this,  the conformal group  can be at most   3-dimensional. Therefore,
if there are theories with larger conformal groups than $SL(2,\bR)$, then necessarily must have additional fields, like vectors or spinors, and possibly must couple to gravity.
As a consequence all linear models admit at most a $SL(2,\bR)$ conformal symmetry.

We have also investigated the conformal properties of 1-dimensional systems with  scalar and vector fields based on the Lagrangian(\ref{actionb}).
We have derived the conditions for such systems to admit a conformal symmetry (\ref{conb})  and present several examples.  The potential of a class of such theories
is again the sum of  a homogeneous function, under the action the homothetic motion, and a term that depends on the homothetic potential.
Examples of such conformal models can exhibit  general gauge groups and  global symmetries. In particular, we have constructed models with arbitrary gauge group
which have the isometries of $AdS_2\times S^3$ and $AdS_2\times S^3\times S^3$ backgrounds as global symmetries. Similar potentials have arisen in the
investigation of matrix models with $SL(2,\bR)$ invariance in \cite{park}.

Gravitational backgrounds that have applications in $AdS_2/CFT_1$ typically preserve some of the spacetime supersymmetry and as a result the dual theories must be superconformal.
The supersymmetric extension of some of the conformal models we have considered here has already been done, see eg \cite{mende}
and \cite{michelson, papadopoulos} for the supersymmetric extension of Calogero model
with harmonic oscillator couplings and that of non-linear conformal theories with homogeneous potentials, respectively, see also \cite{parkb} for matrix models. Conformal linear
models with extended supersymmetry and homogenous potentials have  been reviewed in \cite{olaf}, see also  \cite{toppan}.  
It is straightforward to construct superconformal models with potentials $V=V_0+V_1$ specially those that exhibit a small number of supersymmetries. 
Such supersymmetric extensions can be based on the results of \cite{colesgp, gps} and they will be reported elsewhere.

\vskip 0.5cm
\noindent{\bf Acknowledgements} \vskip 0.1cm
\noindent I would like to specially thank Anton Galajinsky and Jeong-Hyuck Park for
their comments as they have led to significant improvements in the paper. I also like to thank
Evgeny Ivanov and Roman Jackiw for correspondence.
GP is partially supported by the  STFC rolling grant ST/J002798/1.
\vskip 0.5cm

\setcounter{section}{0}\setcounter{equation}{0}




\begin{thebibliography}{99}


 \bibitem{jackiwc} R. Jackiw,  "Introducing Scale Symmetry,"  Physics Today 25, No. 1, 23 (1972).
 
\bibitem{fubini}
  V.~de Alfaro, S.~Fubini and G.~Furlan,
  ``Conformal Invariance in Quantum Mechanics,''
  Nuovo Cim.\ A {\bf 34} (1976) 569.

 

  \bibitem{jackiw}
  R.~Jackiw,
  ``Dynamical Symmetry Of The Magnetic Monopole,''
  Annals Phys.\  {\bf 129} (1980) 183.

  \bibitem{akulov}V.P. Akulov and I.A. Pashnev, ``Quantum
Superconformal Model
in (2,1) Space'', Theor. Math. Phys. {\bf 56} (1983) 862.


\bibitem{fubinib} S. Fubini and E. Rabinovici, ``Superconformal
Quantum Mechanics'',
Nucl. Phys. {\bf B245} (1984) 17.


\bibitem{ivanov} E. Ivanov, S. Krivonos and V. Leviant, ``Geometry of
Conformal Mechanics'', J. Phys. {\bf A22} (1989) 345.


\bibitem{jackiwb}
  R.~Jackiw,
  ``Dynamical Symmetry of the Magnetic Vortex,''
  Annals Phys.\  {\bf 201} (1990) 83.


  \bibitem{wyllard}
N.~Wyllard,
  ``(Super)conformal many body quantum mechanics with extended supersymmetry,''
  J.\ Math.\ Phys.\  {\bf 41} (2000) 2826
  [hep-th/9910160].


\bibitem{michelson}
  J.~Michelson and A.~Strominger,
  ``The Geometry of (super)conformal quantum mechanics,''
  Commun.\ Math.\ Phys.\  {\bf 213} (2000) 1
  [hep-th/9907191].


  \bibitem{papadopoulos}
  G.~Papadopoulos,
  ``Conformal and superconformal mechanics,''
  Class.\ Quant.\ Grav.\  {\bf 17} (2000) 3715
  [hep-th/0002007].


  \bibitem{michelsonb}
  R.~Britto-Pacumio, J.~Michelson, A.~Strominger and A.~Volovich,
  ``Lectures on superconformal quantum mechanics and multiblack hole moduli spaces,''
  hep-th/9911066.


  \bibitem{olaf}
  S.~Fedoruk, E.~Ivanov and O.~Lechtenfeld,
  ``Superconformal Mechanics,''
  J.\ Phys.\ A {\bf 45} (2012) 173001
  [arXiv:1112.1947 [hep-th]].


\bibitem{claus}
P. Claus, M. Derix, R. Kallosh,
J. Kumar, P. Townsend and A. van
Proeyen, ``Black Holes and Superconformal Mechanics,''
 Phys. Rev. Lett.
{\bf 81} (1998) 4553; hep-th/9804177.

\bibitem{jade} J.A. de Azcarraga, J.M. Izquerido,
J.C. Perez Buono and P.K. Townsend,
``Superconformal Mechanics, Black Holes, and
Non-linear Realizations,'' Phys. Rev.
{\bf D59} (1999) 084015; hep-th/9810230.



\bibitem{gibbons}  G. W. Gibbons and P.K. Townsend,
 ``Black Holes and Calogero Models,''
 Phys.Lett. {\bf B454}(1999) 187;
 hep-th/9812034.

 \bibitem{galajinskya}
  A.~Galajinsky,
  ``Particle dynamics on AdS(2) x S**2 background with two-form flux,''
  Phys.\ Rev.\ D {\bf 78} (2008) 044014
  [arXiv:0806.1629 [hep-th]].

   ``Particle dynamics near extreme Kerr throat and supersymmetry,''
  JHEP {\bf 1011} (2010) 126
  [arXiv:1009.2341 [hep-th]].


\bibitem{gibbonsb} G.W.Gibbons \& P.J.Ruback,
``The motion of extreme Reissner-Nordstr\"om black
holes in the low velocity limit,''  Phys. Rev.
Lett. {\bf 57} (1986) 1492.

\bibitem{shiraishi} K. Shiraishi, ``Moduli Space
Metric for Maximally-Charged
Dilaton Black Holes,'' Nucl. Phys. {\bf B402} (1993) 399.

\bibitem{gps} G.W. Gibbons, G. Papadopoulos  and K.S. Stelle,
``HKT and OKT Geometries on Soliton Black Hole Moduli Spaces,''
 Nucl.Phys. {\bf B508} (1997)623; hep-th/9706207.


 \bibitem{micha} J. Michelson and A. Strominger, ``Superconformal
Multi-Black Hole
Quantum Mechanics,'' HUTP-99/A047, hep-th/9908044.

\bibitem{gutpap}  J.~Gutowski and G.~Papadopoulos,
  ``The Dynamics of very special black holes,''
  Phys.\ Lett.\ B {\bf 472} (2000) 45
  [hep-th/9910022].

  ``Moduli spaces for four-dimensional and five-dimensional black holes,''
  Phys.\ Rev.\ D {\bf 62} (2000) 064023
  [hep-th/0002242].

\bibitem{maldacena} J. Maldacena, ``The Large N Limit of
Superconformal Field
Theories and Supergravity,''  Adv. Theor. Math. Phys.
{\bf 2} (1998) 231; hep-th/9711200.


\bibitem{sen}
  A.~Sen,
  ``State Operator Correspondence and Entanglement in $AdS_2/CFT_1$,''
  Entropy {\bf 13} (2011) 1305
  [arXiv:1101.4254 [hep-th]].
  
  \bibitem{santos}
  C.~Chamon, R.~Jackiw, S.~-Y.~Pi and L.~Santos,
  ``Conformal quantum mechanics as the CFT$_1$ dual to AdS$_2$,''
  Phys.\ Lett.\ B {\bf 701} (2011) 503
  [arXiv:1106.0726 [hep-th]].


\bibitem{olafb}
  S.~Fedoruk, E.~Ivanov and O.~Lechtenfeld,
  ``Supersymmetric Calogero models by gauging,''
  Phys.\ Rev.\ D {\bf 79} (2009) 105015
  [arXiv:0812.4276 [hep-th]].


\bibitem{park}
  J.~Erdmenger, J.~-H.~Park and C.~Sochichiu,
  ``Matrix models from D-particle dynamics and the string/black hole transition,''
  Class.\ Quant.\ Grav.\  {\bf 23} (2006) 6873
  [hep-th/0603243].


\bibitem{calogero}
 F.~Calogero,
  ``Solution of the one-dimensional N body problems with quadratic and/or inversely quadratic pair potentials,''
  J.\ Math.\ Phys.\  {\bf 12} (1971) 419.

  \bibitem{galajinsky}
  A.~Galajinsky,
  ``Conformal mechanics in Newton-Hooke spacetime,''
  Nucl.\ Phys.\ B {\bf 832} (2010) 586
  [arXiv:1002.2290 [hep-th]].

\bibitem{new}
G.~ Papadopoulos, to appear

\bibitem{mende}{D.Z. Freedman and P. Mende, ``An
Exactly Solvable N Particle
System in Supersymmetric Quantum Mechanics,''
 Nucl. Phys. {\bf B344} (1990) 317.}

\bibitem{colesgp}
  R.~A.~Coles and G.~Papadopoulos,
  ``The Geometry of the one-dimensional supersymmetric nonlinear sigma models,''
  Class.\ Quant.\ Grav.\  {\bf 7} (1990) 427.

\bibitem{parkb}
  N.~B.~Copland, S.~M.~Ko and J.~-H.~Park,
  ``Superconformal Yang-Mills quantum mechanics and Calogero model with $OSp(N\vert 2,R)$ symmetry,''
  JHEP {\bf 1207} (2012) 076
  [arXiv:1205.3869 [hep-th]].

\bibitem{toppan}
Z.~Kuznetsova and F.~Toppan,
  ``D-module Representations of N=2,4,8 Superconformal Algebras and Their Superconformal Mechanics,''
  J.\ Math.\ Phys.\  {\bf 53} (2012) 043513
  [arXiv:1112.0995 [hep-th]].
  
  S.~Khodaee and F.~Toppan,
  ``Critical scaling dimension of D-module representations of N=4,7,8 Superconformal Algebras and constraints on Superconformal Mechanics,''
  arXiv:1208.3612 [hep-th].


\end{thebibliography}
\end{document}